\documentclass[runningheads]{llncs}
\usepackage{caption}
\usepackage[utf8]{inputenc}
\usepackage[T1]{fontenc}
\usepackage{graphicx,verbatim}
\usepackage{subcaption}
\usepackage{float}
\begin{document}
\title{Beyond the Desktop: XR-Driven Segmentation with Meta Quest 3 and MX Ink}

\titlerunning{Beyond the Desktop: XR-Driven Segmentation}
%
\begin{comment}  %% Removed for anonymized MICCAI 2025 submission
\author{First Author\inst{1}\orcidID{0000-1111-2222-3333} \and
Second Author\inst{2,3}\orcidID{1111-2222-3333-4444} \and
Third Author\inst{3}\orcidID{2222--3333-4444-5555}}
%
\authorrunning{F. Author et al.}
\institute{Princeton University, Princeton NJ 08544, USA \and
Springer Heidelberg, Tiergartenstr. 17, 69121 Heidelberg, Germany
\email{lncs@springer.com}\\
\url{http://www.springer.com/gp/computer-science/lncs} \and
ABC Institute, Rupert-Karls-University Heidelberg, Heidelberg, Germany\\
\email{\{abc,lncs\}@uni-heidelberg.de}}

\end{comment}

\author{Lisle Faray de Paiva\inst{1}\and
        Gijs Luijten\inst{1,2,3} \and
        Ana Sofia Ferreira Santos\inst{1}\and
        Moon Kim\inst{1} \and
        Behrus Puladi\inst{4,5}\and
        Jens Kleesiek\inst{1,6,7,8,9,10} \and
        Jan Egger\inst{1,2,3,6,7,10}}
        
        % leave out streats/no addresses / forschheim germany - check signatures siemens - ask Siemens Healthcare GmbH, 91301 Forchheim, Germany (krueger)

        % Ulf - maastricht      
% Department of Surgery, Maastricht University Medical Centre+, Maastricht, the Netherlands
% OR
% Department of Surgery, NUTRIM School of Nutrition and Translational Research in Metabolism, Maastricht University, Maastricht, The Netherlands

\authorrunning{de Paiva et al.}

\institute{Institute for Artificial Intelligence in Medicine (IKIM), Essen University Hospital (AöR), University of Duisburg-Essen, Essen, Germany \and
           Center for Virtual and Extended Reality in Medicine (ZvRM), University Hospital Essen (AöR), Essen, Germany \and
           Institute of Computer Graphics and Vision (ICG), Graz University of Technology, Graz, Austria \and
           Department of Oral and Maxillofacial Surgery, University Hospital RWTH Aachen, Aachen, Germany \and
           Institute of Medical Informatics, University Hospital RWTH Aachen, Aachen, Germany \and
           Medical Faculty, University of Duisburg-Essen, Essen, Germany \and
           Cancer Research Center Cologne Essen (CCCE), West German Cancer Center, University Hospital Essen (AöR), Essen, Germany \and
           German Cancer Consortium (DKTK), Partner site University Hospital Essen (AöR), Essen, Germany \and
           Technische Universität Dortmund, Fakultät Physik, Dortmund, Germany\and
           Faculty of Computer Science, University of Duisburg-Essen, Essen, Germany
           \\
           \vspace{1em}
           \email{Corresponding authors: Lisle.FaraydePaiva@uk-essen.de, Jan.Egger@uk-essen.de}}

\maketitle              % typeset the header of the contribution

\begin{abstract}
Medical imaging segmentation is essential in clinical settings for diagnosing diseases, planning surgeries, and other procedures. However, manual annotation is a cumbersome and effortful task. To mitigate these aspects, this study implements and evaluates the usability and clinical applicability of an extended reality (XR)-based segmentation tool for anatomical CT scans, using the Meta Quest 3 headset and Logitech MX Ink stylus. We develop an immersive interface enabling real-time interaction with 2D and 3D medical imaging data in a customizable workspace designed to mitigate workflow fragmentation and cognitive demands inherent to conventional manual segmentation tools. The platform combines stylus-driven annotation, mirroring traditional pen-on-paper workflows, with instant 3D volumetric rendering. A user study with a public craniofacial CT dataset demonstrated the tool's foundational viability, achieving a System Usability Scale (SUS) score of 66, within the expected range for medical applications. Participants highlighted the system's intuitive controls (scoring 4.1/5 for self-descriptiveness on ISONORM metrics) and spatial interaction design, with qualitative feedback highlighting strengths in hybrid 2D/3D navigation and realistic stylus ergonomics. While users identified opportunities to enhance task-specific precision and error management, the platform's core workflow enabled dynamic slice adjustment, reducing cognitive load compared to desktop tools. Results position the XR-stylus paradigm as a promising foundation for immersive segmentation tools, with iterative refinements targeting haptic feedback calibration and workflow personalization to advance adoption in preoperative planning. 

\keywords{Extended reality \and Image segmentation \and Medical imaging.}
% Authors must provide keywords and are not allowed to remove this Keyword section.

\end{abstract}

\section{Introduction}
Three-dimensional segmentation of anatomical structures in medical imaging is critical for applications ranging from preoperative planning to personalized treatment design \cite{ma2024segment,johnsonchris2022review}. This process involves identifying and delineating organs, lesions, and tissues across imaging modalities, supporting clinical workflows such as disease diagnosis, surgical navigation, and therapy monitoring \cite{ma2024segment}. However, conventional manual segmentation workflows remain labor-intensive \cite{ma2024segment,johnsonchris2022review}. These limitations can hinder the integration of segmentation tasks into time-sensitive clinical routines such as preoperative planning \cite{ma2024segment,johnsonchris2022review}. 

Existing segmentation tools often exacerbate these challenges through ergonomically suboptimal interfaces and workflow fragmentation through additional software requirements \cite{prabhu2005ergonomics,multi2011human}. Clinicians must mentally reconcile 2D cross-sectional views with 3D reconstructions and navigate complex software menus, increasing cognitive load and hampering efficiency. Furthermore, traditional input devices like mice poorly approximate the natural dexterity required for precise anatomical delineation \cite{multi2011human,yarmand2024enhancing}. 

Extended reality (XR) technologies address these limitations by enabling direct 3D interaction with medical imaging data, unifying 2D slice annotation, and volumetric manipulation in a spatial workspace that mirrors real-world clinical workflows \cite{kukla2023extended,andrews2019extended}. Recent advances in XR segmentation tools have demonstrated their potential to improve medical image segmentation accuracy and user interaction, such as AR-ViSeM introduces an Augmented reality (AR) tool for visualization, segmentation, and interactive mask editing, allowing users to manipulate medical imaging data in real-time without the need for complex installations \cite{charalampidisar}. Similarly, graph-based segmentation in XR has been applied for multi-organ 3D reconstruction, improving structural detail preservation and enhancing the immersive learning experience for medical students \cite{nagendra2024multi}. Gruber et al. \cite{gruber2024accuracy} analyze the XR application Elucis, developed by Realize Medical Inc., which employs the HTC Vive Pro headset alongside the VR stylus Ink Pilot Edition and the VR Ink Drawing Mat. While it features six Degrees-of-Freedom (6DoF) spatial tracking, it requires externally mounted sensors for operation.

To bridge this gap, we develop an XR-based segmentation platform combining the Meta Quest 3 headset and the Logitech MX Ink stylus. By leveraging the MX Ink integration with Meta Quest's built-in 6DoF spatial tracking, the proposed application enables direct manipulation of 2D slices and 3D reconstructions in a unified workspace, supporting workflow acceleration through pre-annotated mask integration. To evaluate the clinical viability of our XR platform, we conducted a user study where participants were tasked with segmenting anatomical structures in craniofacial CT scans. The study employs standardized usability metrics, including the System Usability Scale (SUS) and an ISO 9241-110 compliant questionnaire, to quantify ergonomic integration, workflow efficiency, and the MX Ink stylus's suitability for medical applications, an area lacking prior research. These metrics are contextualized against industry benchmarks.

\section{Methodology}

\subsection{System Design and Implementation}
\subsubsection{Development Framework:}
The segmentation tool was developed in Unity 2022.3.50f1 using the Meta XR SDK and the OpenXR plugin to support cross-platform deployment. The application utilizes the Meta Quest 3 standalone headset, leveraging its Snapdragon XR2 Gen 2 processor for real-time mesh rendering. User interaction is enabled through a paired Logitech MX Ink stylus for intuitive annotation, complemented by a Quest controller for User Interface (UI) navigation.

Medical imaging data is integrated via TiffLib, a custom TIFF parsing module that directly imports 3D CT scans as TIFF stacks (converted from DICOM) and optional segmentation masks (e.g., AI-generated .tiff files). This ensures compatibility with clinical pipelines while maintaining the original voxel spacing.

\subsubsection{Spatial Interaction Pipeline:}

The application begins by retrieving DICOM-derived TIFF stacks. These stacks are parsed using LibTiff and converted into 8-bit textures at a fixed resolution (512 × 512 pixels). A preloading routine buffers a subset of image slices into GPU memory, ensuring fast and responsive navigation and improving loading time. Users can seamlessly switch between 2D slice viewing and 3D surface exploration, with features such as overlay compositing, blending segmentation masks onto the original images, and pipelines enabling the batch export of segmentation masks.

\begin{figure}[ht]
    \centering
    \captionsetup[subfigure]{font=small,labelfont=bf,labelsep=period}
     \begin{subfigure}[b]{0.35\textwidth}
          \centering
          \includegraphics[width=\textwidth]{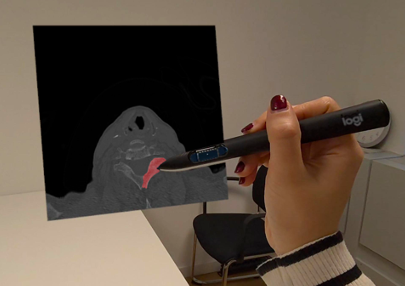}
          \caption{User's perspective during freehand annotation (virtual canvas).}
          \label{fig:stand_draw}
      \end{subfigure}
      \hspace{1cm}
      \begin{subfigure}[b]{0.35\textwidth}
          \centering
          \includegraphics[width=\textwidth]{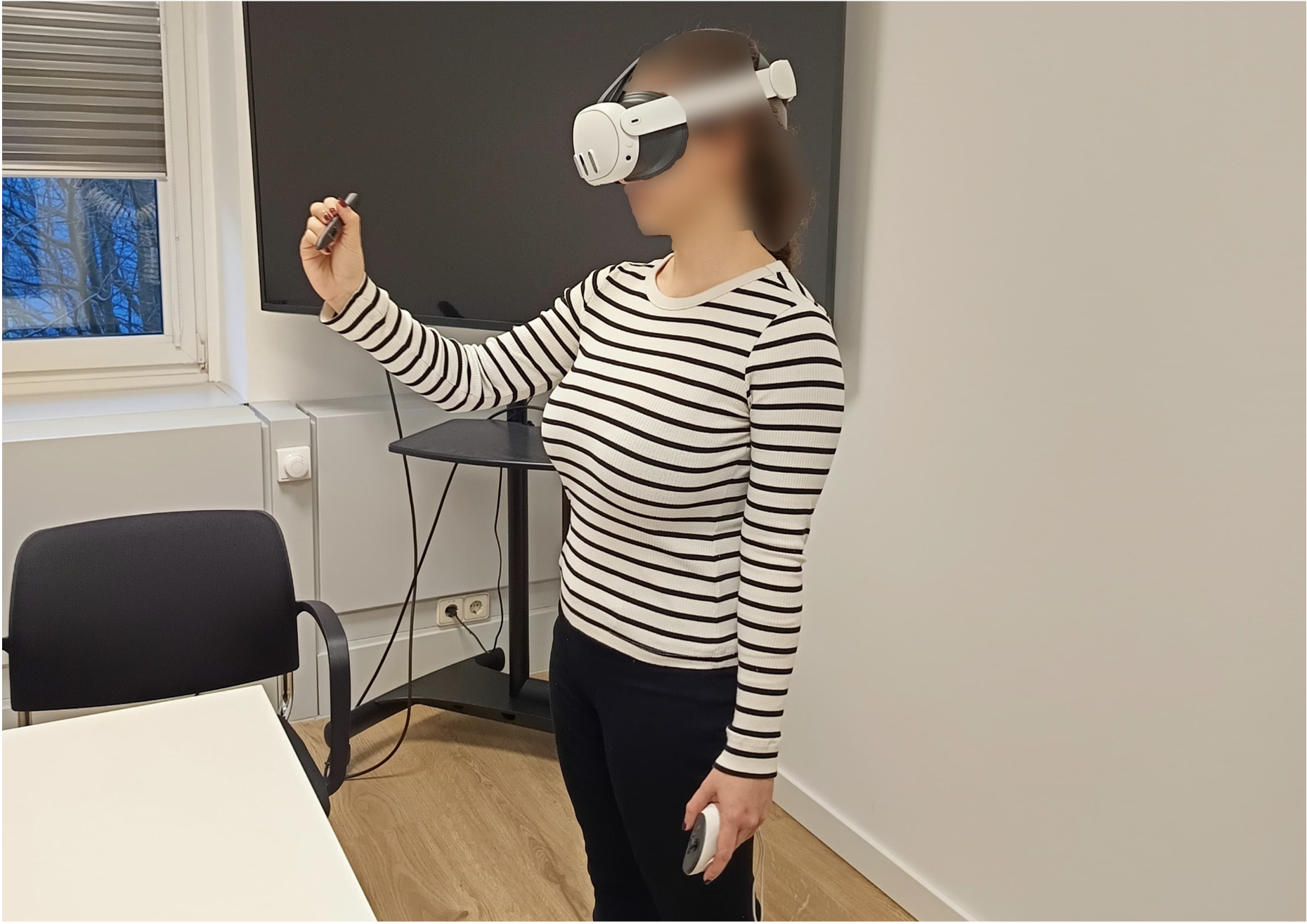}
          \caption{Real-world view: Standing posture \& stylus positioning.}
          \label{fig:stand_real}
      \end{subfigure}

    \hspace{-1cm}
      
      % Second row of subfigures
      \begin{subfigure}[b]{0.35\textwidth}
          \centering
          \includegraphics[width=\textwidth]{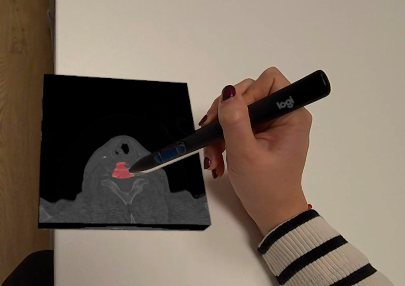}
          \caption{Surface-aligned annotation (user's perspective).}
          \label{fig:sit_draw}
      \end{subfigure}
      \hspace{1cm}
      \begin{subfigure}[b]{0.35\textwidth}
          \centering
          \includegraphics[width=\textwidth]{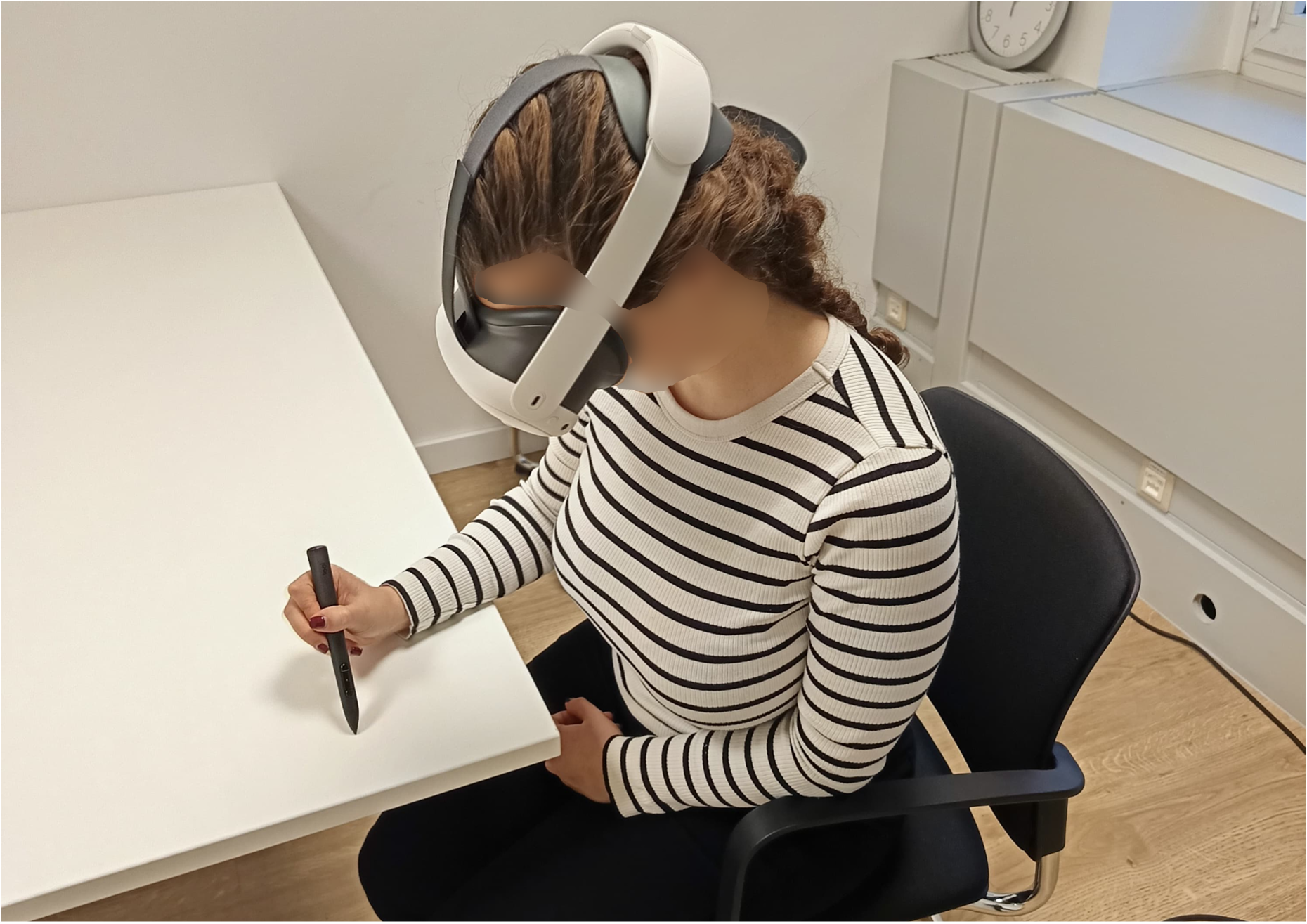}
          \caption{Physical environment integration (real-world view).}
          \label{fig:sit_real}
      \end{subfigure}
      
    \caption[XR interface workflow]{Segmentation interface demonstrating: (a,c) First-person annotation views with virtual canvas; (b,d) Corresponding real-world perspectives showing ergonomic positioning. Privacy-preserving blur applied to real-world images.}
    \label{fig:segmentation}
\end{figure}

To replicate the tactile feeling of pen-on-paper, the XR interface integrates the Logitech MX Ink stylus, enabling spatially aligned segmentation directly on 3D medical images, see Figure \ref{fig:segmentation}, by guaranteeing the positional equivalence between the drawn pixels and the pixel coordinates in the medical image. The pixel matching is established by translating the stylus positioning into the medical image's coordinates system via a UV mapping process that employs raycasting from the stylus tip to an dynamic 3D canvas containing the medical image. For a continuous contour rendering, drawn points are linearly interpolated, minimizing tracking jitter and gaps. Additionally, the system leverages configurable user input to adjust brush and eraser sizes dynamically. At the same time, haptic pulses confirm contact with the canvas, simulating contact between the stylus and the virtual canvas. Additive and subtractive modes allow clinicians to refine segmentations iteratively, preserving the intuitive workflow of traditional annotation while leveraging 3D spatial context.

\begin{figure}
    \centering
    \begin{subfigure}[b]{0.35\textwidth}
          \centering
          \includegraphics[width=\textwidth]{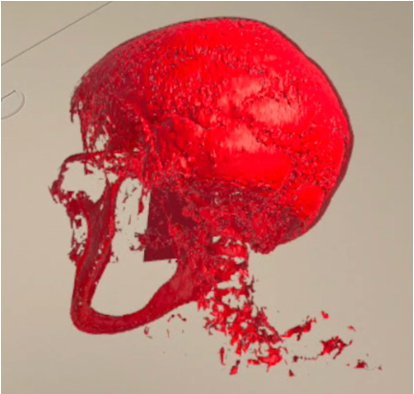}
          \caption{Lateral view of the volumetric skull reconstruction.}
          \label{fig:3d_lateral}
      \end{subfigure}
      \hspace{1cm}
      \begin{subfigure}[b]{0.35\textwidth}
          \centering
          \includegraphics[width=\textwidth]{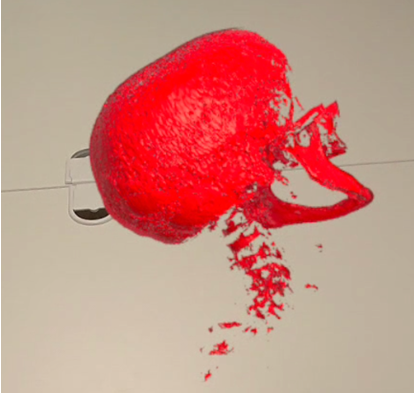}
          \caption{Oblique view of the volumetric skull reconstruction.}
          \label{fig:3d_oblique}
      \end{subfigure}
    \caption{Real-time volumetric rendering of segmented head CT shown in red (back of skull and mandible).}
    \label{fig:3dvolume}
\end{figure}

As seen in Figure \ref{fig:3dvolume}, volumetric visualization begins by generating an anatomically accurate iso-surface from either the segmentation masks or the whole scan based on Hounsfield units using the Marching Cubes algorithm \cite{lorensen1998marching}. First, the slice stacks are converted to a 3D density grid, preserving the original scan's voxel dimensions (0.3 x 0.3 x 0.5 $mm^{3}$ for the data used in this study). Each grid cell stores a density value thresholded against an iso-level ($\iota = 0.5$) to distinguish foreground anatomy. The grid is then processed cube-by-cube using marching cubes, generating vertices and triangles at density transitions. The algorithm employs dynamic yield intervals (every 1000th cube) to maintain interactive performance to prevent UI thread blocking, enabling asynchronous mesh updates during computation. To meet real-time performance requirements, the generated mesh undergoes further optimization through mesh simplification by a quadric edge collapse decimation, reducing triangle count by 70\%-80\% while preserving critical anatomical features, vertex deduplication, and adaptive resolution via downsampling based on viewport proximity. These optimizations ensure the 3D volume can be updated interactively (with latencies below 200ms) and sustained 72 FPS rendering on Meta Quest 3 hardware.

\subsection{Evaluation Protocol}

The study involved ten participants, comprising medical students and practicing clinicians, recruited to assess the workflow viability of the XR-stylus interface. Participants were asked to segment anatomical structures in head CT scans from the public Open-Full-Jaw dataset \cite{wallner2019computed}, after which they completed a series of questionnaires before any discussion sessions.

The Open-Full-Jaw Dataset is a repository of 3D volumes of craniofacial anatomy comprising high-resolution clinical CT scans of complete, teeth-free human mandibles derived from ten patient cases. These scans capture anatomically realistic craniofacial structures with complex bone boundaries, offering a standardized testbed for evaluating segmentation tools. While the dataset includes expert-validated labels and meshes for algorithm development, only the CT volumes were used to ensure participants focused on the XR-stylus interface's usability over segmentation accuracy.

Usability was evaluated through a mixed-methods approach. First, the System Usability Scale (SUS) quantified overall software usability. The SUS is a validated 10-item Likert scale instrument with positive and negative statements to capture a holistic view of user experience \cite{brooke1996sus}. This instrument enabled us to obtain a single usability score reflecting ease of use and overall satisfaction with the software.

In addition to the SUS, stylus ergonomics, and interaction fidelity was assessed using a 10-item questionnaire with nine items compliant with ISO 9241-110 usability principles (suitability for task, self-descriptiveness, controllability), supplemented by an overall satisfaction metric \cite{heinold2025usability}. Specifically, Q1 addressed suitability for the task, Q2 self-descriptiveness, Q3 controllability, Q4 conformity with user expectations, Q5 and Q6 error tolerance, Q7 suitability for individualization, Q8 and Q9 suitability for learning, and Q10 measured overall satisfaction \cite{heimgartner2014human}. Internal consistency of the adapted ISO instrument was confirmed via Cronbach's alpha ($\alpha = 0.89$). 

In addition to these standardized instruments, demographic information and open-ended feedback were collected through semi-structured interviews. These interviews were designed to capture qualitative insights regarding hardware-software integration and the system's clinical applicability.

The evaluation protocol was executed in three phases: a 10-minute training session introducing XR navigation and stylus controls, a 30-minute task execution period during which participants employed a think-aloud protocol, and a post-task debrief combining the SUS and ISO questionnaires with open-ended interviews. This comprehensive approach ensured that objective measures and subjective insights were captured, supporting a robust analysis of the system's usability in a clinical context.

\section{Results}

A total of ten participants took part in the study, consisting of medical students and practicing clinicians. The demographic breakdown revealed that participants' ages ranged from 20 to 27 years, with a gender distribution of 60\% male and 40\% female. The majority (70\%) of participants did not wear glasses. Of those who wore glasses, 50\% had a mild prescription strength between -1.0 and -2, while the remaining participants had no specified prescription. Regarding previous experience with head-mounted displays (HMDs), 40\% of participants had prior experience with optical and video pass-through HMDs, while 60\% had no such experience.

\begin{figure}
    \centering
    \includegraphics[width=0.7\linewidth]{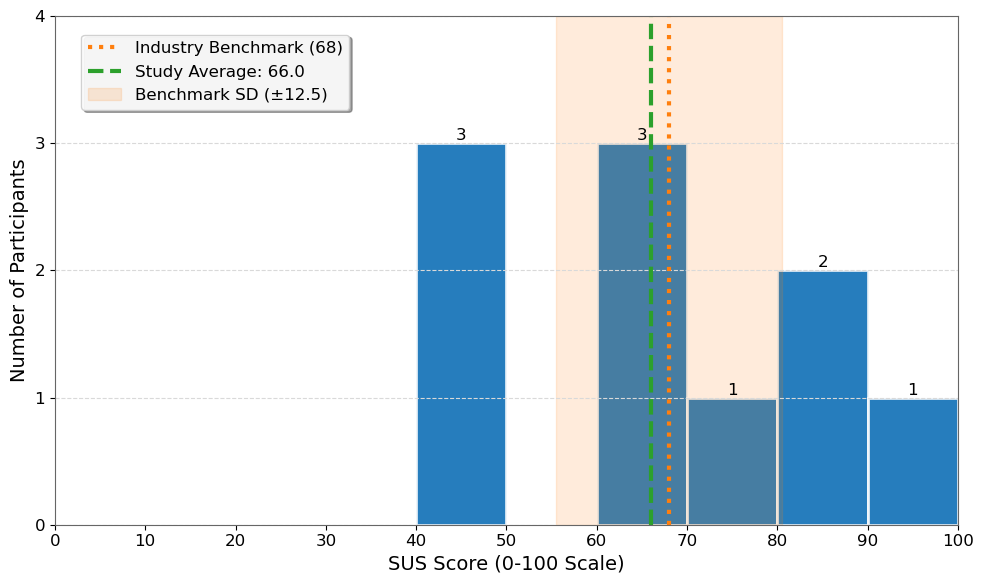}
    \caption{Distribution of System Usability Scale (SUS) scores (0–100) across participants. The study’s average (66.0, green dashed line) aligns closely with the industry benchmark (68, orange dashed line) and lies within the standard deviation range (±12.5, shaded region). Benchmark source: Hyzy et al. (2022).}
    \label{fig:sus}
\end{figure}

The SUS questionnaire yielded an average score of 66.0 across participants. Individual scores varied, spanning a range that included both below and above the commonly referenced benchmark of 68 for industry usability studies and digital health applications \cite{lewis2018item,hyzy2022system}. Three participants scored under 60, four scored between 60 and 70, two were in the 70-80 range, and one exceeded 80 as showcased in Figure \ref{fig:sus}.

\begin{figure}[h]
    \centering
    \includegraphics[width=0.7\linewidth]{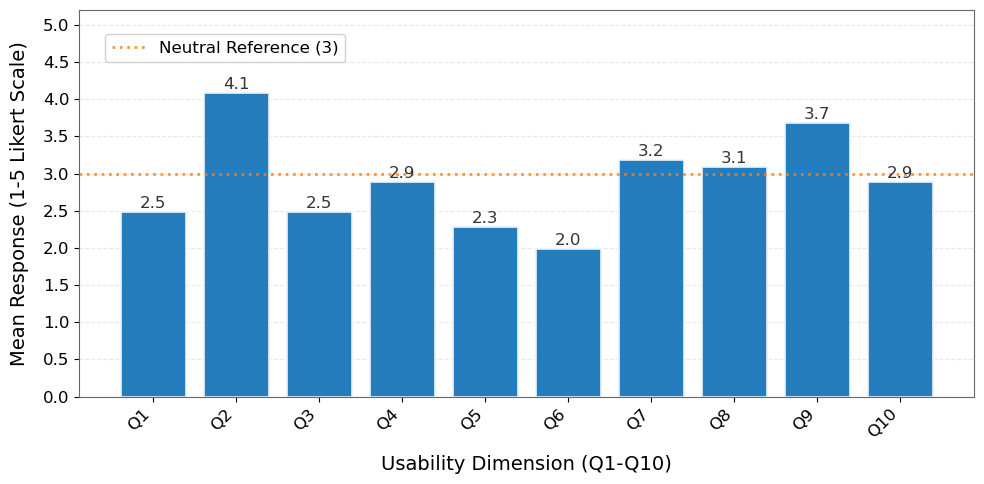}
    \caption{Mean responses to the 10-item ISO 9241-110 S compliant questionnaire. Items above the neutral reference (dashed line at 3.0) indicate relatively higher perceived usability in those dimensions, while items below three highlight areas for improvement.}
    \label{fig:iso}
\end{figure}

The custom ISO-compliant questionnaire produced mean item ratings between 2.5 and 4.1 on a 1(strongly disagree) - 5 (strongly agree) Likert scale (Figure \ref{fig:iso}) following the SUS pattern. Questions relating to clarity of controls and operational cues (Q2) and ease of learning (Q8) showed higher mean values ($\geq 3.7$), while other items (e.g., Q1, Q3) averaged near or below the neutral reference of 3.0. The overall internal consistency of this 10-item questionnaire, as measured by Cronbach's alpha, was 0.89.

Participants cited software stability, interface simplicity, and "real-feel" annotation simulating drawing on paper as strengths during the open-ended feedback. Reported weaknesses included inconsistent pen tracking near the edge of the canvas and limited error-correction options in case of drawing mistakes. Suggestions for improvement ranged from adding a dedicated "undo function" and refining stylus interactions to introducing built-in AI-assisted segmentation features.

\section{Discussion}

The overall usability of the XR application, as measured by the SUS, produced a mean score of 66, which is slightly below the conventional benchmark of 68 for industrial usability studies \cite{lewis2018item} but remains well within the accepted standard deviation of 12.5 for digital health applications \cite{hyzy2022system}. Given that this work is a prototype, the SUS score is consistent with early-stage systems where prototypical behavior and functionality gaps are expected. This finding suggests that, while the current system may benefit from further refinement, its usability is within a range considered acceptable for preliminary evaluations in clinical contexts.

A recurring strength highlighted by participants was the MX Ink stylus's ability to simulate real-world drawing workflows, which directly reduced cognitive effort during segmentation. Qualitative feedback highlighted the stylus's ergonomic design, with one participant noting it "\textit{felt like annotating on paper, but in 3D space.}". This aligns with the ISO questionnaire results, where the stylus scored highest on self-descriptiveness (4.1/5), indicating that its intuitive, pen-like interaction lowered the learning curve for clinicians accustomed to traditional annotation tools. By mirroring the tactile precision of pen-on-paper feeling while enabling direct 3D manipulation, the stylus bridged the gap between conventional methods and immersive technology, a critical factor for clinical adoption. 

The Meta Quest 3's standalone design was also recognized for its portability and spatial customization, allowing participants to adapt their workspace dynamically. Unlike desktop-based systems constrained to external monitors and wired peripherals, participants could reposition the virtual 3D canvas onto real-world surfaces (e.g., desk, wall) as shown in Figures \ref{fig:stand_real} and \ref{fig:sit_real}, aligning annotation tasks with ergonomic preferences. Participants also highlighted the wireless setup's ability to reduce physical clutter and support on-demand annotation in diverse clinical environments, such as transitioning from radiology workstations to collaborative conference rooms without workflow interruption.

Looking beyond segmentation, participants identified medical education as a promising application. The system's hybrid 2D/3D interface could serve as a scaffold for teaching anatomical spatial reasoning, particularly for complex structures like the mandible, where traditional 2D atlases struggle to convey depth relationships. Future iterations could leverage the Meta Quest's native multiuser capabilities to enable collaborative training sessions, where instructors and trainees interact with the same 3D reconstruction in real time. Such a feature would allow educators to demonstrate segmentation techniques spatially or guide students through anatomical landmarks in a shared virtual environment, a significant advance over current screen-based e-learning tools.

However, technical limitations persist with occasional stylus tracking inaccuracies during fine tasks (e.g., delineating thin mandibular canals) align with the ISO controllability score (3.0/5). The headset’s processing constraints also highlight needs for optimized mesh decimation or hybrid cloud-edge architectures.

\section{Conclusion}

This study demonstrates that XR-stylus systems like the Meta Quest 3/MX Ink platform can meaningfully address persistent challenges in medical segmentation workflows. With a SUS score of 66 and strong ISO ratings for stylus intuitiveness (4.1/5 self-descriptiveness), the prototype successfully bridges conventional 2D annotation practices with immersive 3D interaction. Clinicians particularly valued the system's ability to reduce cognitive load through spatial workspace unification - enabling simultaneous manipulation of cross-sectional slices and volumetric reconstructions without traditional desktop workflow fragmentation.

While tracking inconsistencies during fine motor tasks (controllability score: 3.0/5) and standalone hardware limitations highlight areas for refinement, the platform's core achievements, natural stylus ergonomics, and hybrid 2D/3D navigation establish XR as a viable evolution pathway for clinical segmentation tools. Future work should prioritize stylus precision refinement and AI integration workflows to enhance precision and swiftness while leveraging the Quest 3's multiuser capabilities for collaborative annotation sessions. These advancements could position XR systems as alternatives to desktop tools and reimagine how clinicians interact with medical imaging data.

\section{Acknowledgments}
Ana Sofia Ferreira Santos receives funding by the European Union under Grant Agreement 101168715. Views and opinions expressed are however those of the author(s) only and do not necessarily reflect those of the European Union. Neither the European Union nor the granting authority can be held responsible for them.
This work was supported by the REACT-EU project KITE (grant number: EFRE-0801977, Plattform für KI-Translation Essen, https://kite.ikim.nrw/).

\bibliographystyle{splncs04}
\bibliography{mybibliography}

\end{document}